\documentclass[aps,pra,superscriptaddress,amsmath,amssymb,preprintnumbers,floatfix,showpacs,11pt]{revtex4}

\usepackage{amssymb}
\usepackage{epsfig}

\begin{document}
\title{Revival-collapse phenomenon in the quadrature squeezing
of the multiphoton intensity-dependent Jaynes-Cummings model }
\author{ Faisal A. A. El-Orany }

 \affiliation{ Department of Mathematics  and Computer Science,
Faculty of Science, Suez Canal University,
 Ismailia, Egypt}

\date{\today}

\begin{abstract}
For multiphoton intensity-dependent Jaynes-Cummings model (JCM),
which is described by two-level atom interacting with a radiation
field, we prove that there is a relationship  between the atomic
inversion and the quadrature squeezing. We give the required
condition to obtain best information from this relation. Also we
show that this relation is only sensitive to large values of the
detuning parameter. Furthermore, we discuss briefly such relation
for the off-resonance standard JCM.
\end{abstract}

   \maketitle

\section{Introduction}
The interaction between the radiation field and matter (, i.e., atom),
which is usually called Jaynes-Cummings model
 (JCM) \cite{jay1}, is one of the basic systems in quantum optics.
 The JCM system has become experimentally realizable with
the Rydberg atoms in high-$Q$ microwave cavities (, e.g., see \cite{remp}).
Various extensions for JCM
have been performed and investigated in  greater details such as multiphoton \cite{fa},
intensity-dependent
 \cite{buck}, multimode, e.g.,
\cite{fas3,{fas1},{fas2}}, multilevel
atoms \cite{multil} and multiatom interactions \cite{multia}.
 In the framework of rotating-wave approximation
 the JCM has been exactly solved and many interesting features
  have been remarked, such as
sub-Poissonian statistics \cite{kim2}, quadrature squeezing \cite{meys}
 and revival-collapse phenomenon (RCP) in the Rabi oscillation \cite{eber}.
 Furthermore, JCM has been used for generating nonclassical
states via the conditional measurement technique \cite{cm1}.
As well known the most important phenomenon  is the RCP in the
evolution of the atomic inversion $\langle\hat{\sigma}_{z}(T)\rangle$
where both the collapse and the revival have quantum origins, the former is a
 less clear-cut quantum effect: the collapse appears even for the
 stochastic classical field, while the revival finds its origin in the
 granularity of the quantized radiation field. The revivals result from
 the beating of all nearest-neighbor Rabi oscillators and occur when the
 $\bar{n}$th and $(\bar{n}+1)$th oscillators become in phase \cite{gora}.
Moreover,  the envelope of each revival is  readout
of the photon distribution, in particular, for the states whose photon-number
 distributions are slowly varying \cite{fleisch}.
For two-level atom in the excited (or ground) state
interacting resonantly with the radiation fields in  coherent states with
strong initial intensities--as far as
we know--there are three different forms for the shape of the RCP
 based on the type of the Hamiltonian model:
 (i) For single-photon JCM \cite{eber}
there is an initial
collapse of these oscillations followed by regular revivals that slowly
become broader and eventually overlap. In this case
 the revival pattern
has a Gaussian shape. (ii) For two-photon JCM \cite{sukm}
 and/or single-photon intensity-dependent JCM (IJCM) \cite{buck} the revival patterns are periodic (with period $\pi$),
 compact and systematic.
(iii) For single-photon two-mode JCM
the revival series is compact, each revival is followed by secondary
revival and the locations
of the revival patterns in the time interaction "domain" are independent of the
intensities of the initial fields.
Finally, observation of RCP has been
performed using the one-atom mazer \cite{remp}, which is more sophisticated than the
dynamics of the JCM.
 Moreover, schemes for observing RCP via homodyne
detection \cite{faisal2}, photon counting experiment and homodyne
tomography \cite{fas1} have been reported.

Recently we have developed a new technique showing how within the
single-mode multiphoton JCM the RCP of the atomic inversion of the
 single-photon JCM \cite{faisal2} and two-photon JCM \cite{fais2} can
 be manifested in the evolution of the quadrature squeezing of
the radiation field. Also the technique has been applied to
 two-mode single-photon JCM \cite{fais3} and to higher-order squeezing \cite{fais1}.
The technique is based on two approaches: natural and numerical-simulation approaches.
For the former there
is a class of states whose squeezing factors  can directly
include information
on the corresponding atomic inversion,
however, for the latter
the evolution of the quadrature squeezing of
the three-photon JCM  reflects the RCP involved in
the $\langle\hat{\sigma}_{z}(T)\rangle$ of the single-photon JCM for the same
initial field state.
In this paper we apply this technique for the multiphoton
IJCM. For such system we derive the rescaled squeezing factor, which can
give
information on the atomic inversion of the standard, i.e. single-photon
transition, IJCM. Also we discuss the required condition for
obtaining best information from the quadrature squeezing about the atomic
inversion and investigate
 the influence of the detuning parameter on such relation.
Moreover, we investigate these two issues for the standard JCM
(SJCM) since they have not been discussed yet.
These results and those in
\cite{faisal2,{fais2},{fais1}}
show that the RCP occurred in
$\langle\hat{\sigma}_{z}(T)\rangle$ can be detected using
techniques similar to those used for quadrature squeezing, e.g.
 homodyne detector \cite{homo}, nonlinear
homodyne detector \cite{wilk} and multiport homodyne detector \cite{walk}.
It is worth mentioning that in cavity QED, the homodyne detector
technique has been applied to the single Rydberg atom and one-photon field
for studying  the field phase evolution of the regular JCM \cite{haroc}.
The final remark the generation of nonclassical squeezed light in the
IJCM has been reported in \cite{hil}.

The paper is prepared in the following order.
In section 2 we give the basic relations and equations including the
Hamiltonian model, the definition of quadrature squeezing and discuss
the influence of the detuning parameter on the evolution of the
$\langle\hat{\sigma}_{z}(T)\rangle$.
In sections 3 and 4 we investigate the occurrence
of RCP in the squeezing factors for
the resonance and off-resonance cases, respectively.
In section 5 the main
conclusions are summarized.

%%%%%%%%%%%%%%%%%%%%%%%%%%%%%%%%%%%%%%%%%%%%%
\section{Basic relations and equations}
%%%%%%%%%%%%%%%%%%%%%%%%%%%%%%%%%%%%%%%%%%%%%%
In this section we give the basic relations and equations, which mainly be used
in the paper. Precisely, we write down the Hamiltonian model of the system,
 its wave function and the definition of quadrature squeezing.
Additionally we shed briefly the light
on the influence of detuning parameter on the evolution of
the atomic inversion.

The Hamiltonian controlling the IJCM in the rotating wave
approximation is \cite{avia}:

%%%%%%%%%%%%%%%%%%%%%%%%%%%%%%%%%%%%%%%%%%%%%%%%%%%%%%%%%%%%%%%%%%%%%%%%
\begin{equation}
\frac{\hat{H}}{\hbar}=
\omega\hat{a}^{\dagger}\hat{a}+
\frac{1}{2}\omega_{a}\hat{\sigma}_{z}+
\lambda (\hat{R}\hat{\sigma}_{+} +
\hat{R}^{\dagger}\hat{\sigma}_{-}),
 \label{6}
\end{equation}
%%%%%%%%%%%%%%%%%%%%%%%%%%%%%%%%%%%%%%%%%%%%%%%%%%%%%%%%%%%%%%%%%%%%%%%%%%%
where
\begin{equation}
\hat{R}=\hat{a}^m f(\hat{n}), \qquad \hat{n}=\hat{a}^{\dagger}\hat{a}
\label{me}
\end{equation}
and $f(.)$ is a function in the mean-photon number, which is restricted
to $f(\hat{n})=\sqrt{\hat{n}}$
for IJCM and $f(.)=1$ for SJCM.
Also $\hat{\sigma}_{\pm}$ and $\hat{\sigma}_{z}$ are the Pauli spin
operators;
$\omega$ and $\omega_{a}$ are the frequencies of
the cavity mode $\hat{a}$ and the atom, respectively;  $\lambda$
 is the atom-field coupling
constant and $m$ is the transition parameter.

For obtaining the dynamical state of (\ref{6}) we define
 two  operators $\hat{F}_{1}$ and $\hat{F}_{2}$ as
\begin{equation}
\hat{F}_{1}=
\omega (\hat{a}^{\dagger}\hat{a}+
\frac{m}{2}\hat{\sigma}_{z}),\qquad
\hat{F}_{2}=\frac{\Delta}{2}\hat{\sigma}_{z}+
\lambda (\hat{R}\hat{\sigma}_{+} +
\hat{R}^{\dagger}\hat{\sigma}_{-}), \label{7}
\end{equation}
where $\Delta =\omega_{a}-m\omega$ is the detuning parameter.
It is easy to prove that $\hat{F}_{1}$ and $\hat{F}_{2}$ are constants of motion.
Basically, we consider the field and atom are initially prepared in coherent
$|\alpha\rangle$ and excited atomic $|+\rangle$ states, respectively.
In the interaction picture the dynamical state of the system can be
evaluated as:
\begin{eqnarray}
\begin{array}{lr}
|\psi (T)\rangle=\exp(-it\hat{F}_{2})|+,\alpha\rangle
\\
\\
=\sum\limits_{n=0}^{\infty}C_n\left[G_1(n,m,T)|+,n\rangle+
G_2(n,m,T)|-,n+m\rangle\right],\label{8}
\end{array}
\end{eqnarray}
where
\begin{eqnarray}
\begin{array}{lr} T=\lambda t,\quad \eta=\frac{\Delta}{\lambda},
\qquad C_n=\exp(-\frac{1}{2}|\alpha|^2)
\frac{\alpha^n}{\sqrt{n!}},\qquad
\gamma_{n,m}=\sqrt{\frac{\eta^2}{4}+\frac{(n+m)!}{n!}f^{2}(n+m)},\\
\\
G_1(n,m,T)=\cos (T\gamma_{n,m})-i\frac{\eta}{2\gamma_{n,m}}
\sin (T\gamma_{n,m}),\\
\\
G_2(n,m,T)=-i\sqrt{\frac{(n+m)!}{n!}}f(n+m)
\frac{\sin (T\gamma_{n,m})}{\gamma_{n,m}} \label{9}
\end{array}
\end{eqnarray}
and $|-\rangle$ denotes ground atomic state.
Throughout the paper we consider $\alpha$ to be real.
The atomic inversion related to (\ref{8}) is
\begin{equation}
\langle\hat{\sigma}_{z}(T)\rangle=\sum\limits_{n=0}^{\infty}P(n)
\Bigl[ \frac{\frac{\eta^2}{4}}{\gamma_{n,m}^{2}}+
\frac{\frac{(n+m)!}{n!}f^2(n+m)}{\gamma_{n,m}^{2}}\cos (2T
\gamma_{n,m})\Bigr],
\label{da7}
\end{equation}
where $P(n)=|C_n|^2$.
%%%%%%%%%%%%%%%%%%%%%%%%%%%%%%%%%%%%%%%%%%%%%%%%%%%%%%%%%%%%%%%
\begin{figure}
{\includegraphics[width=8cm]{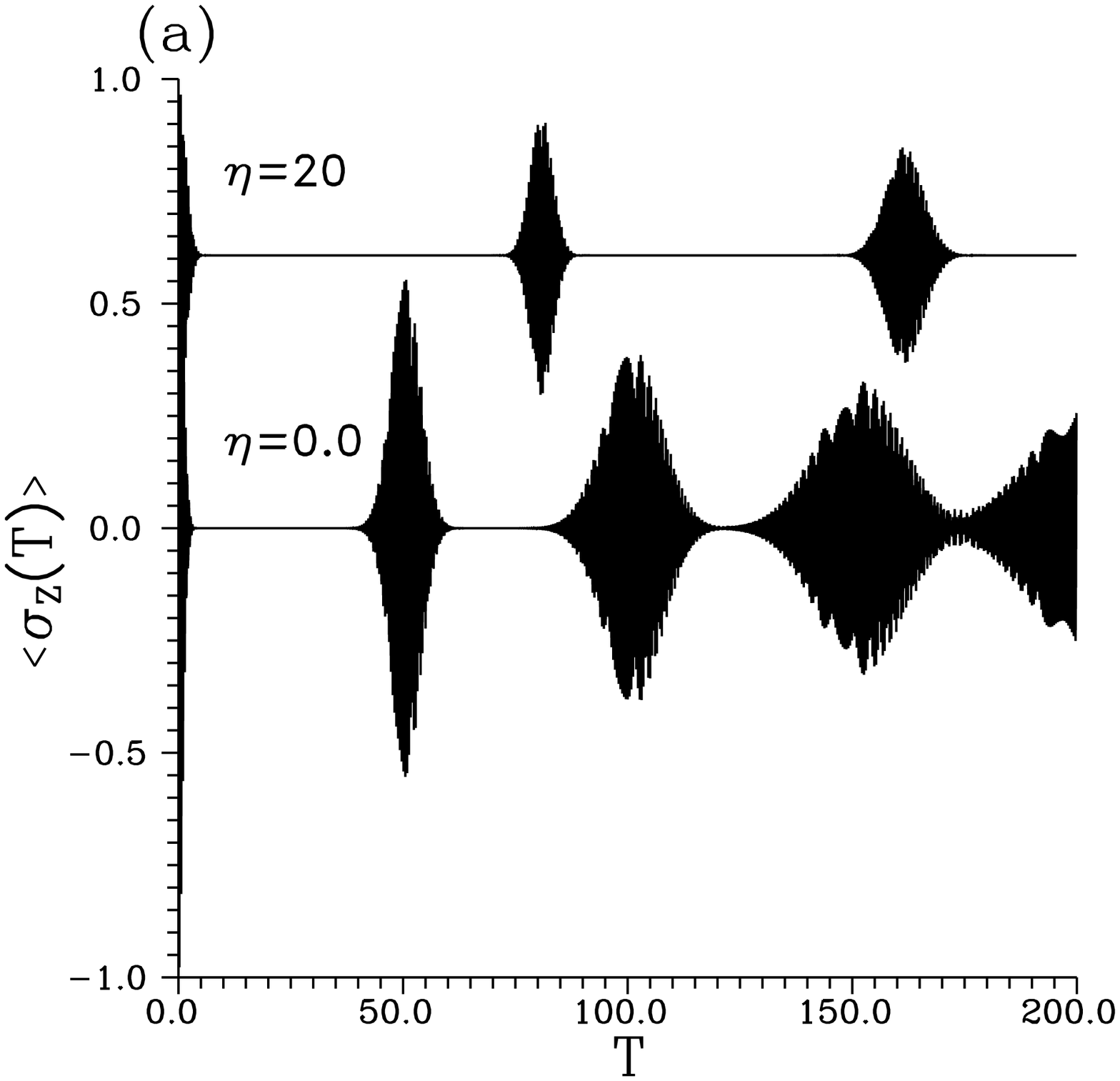}}
{\includegraphics[width=8cm]{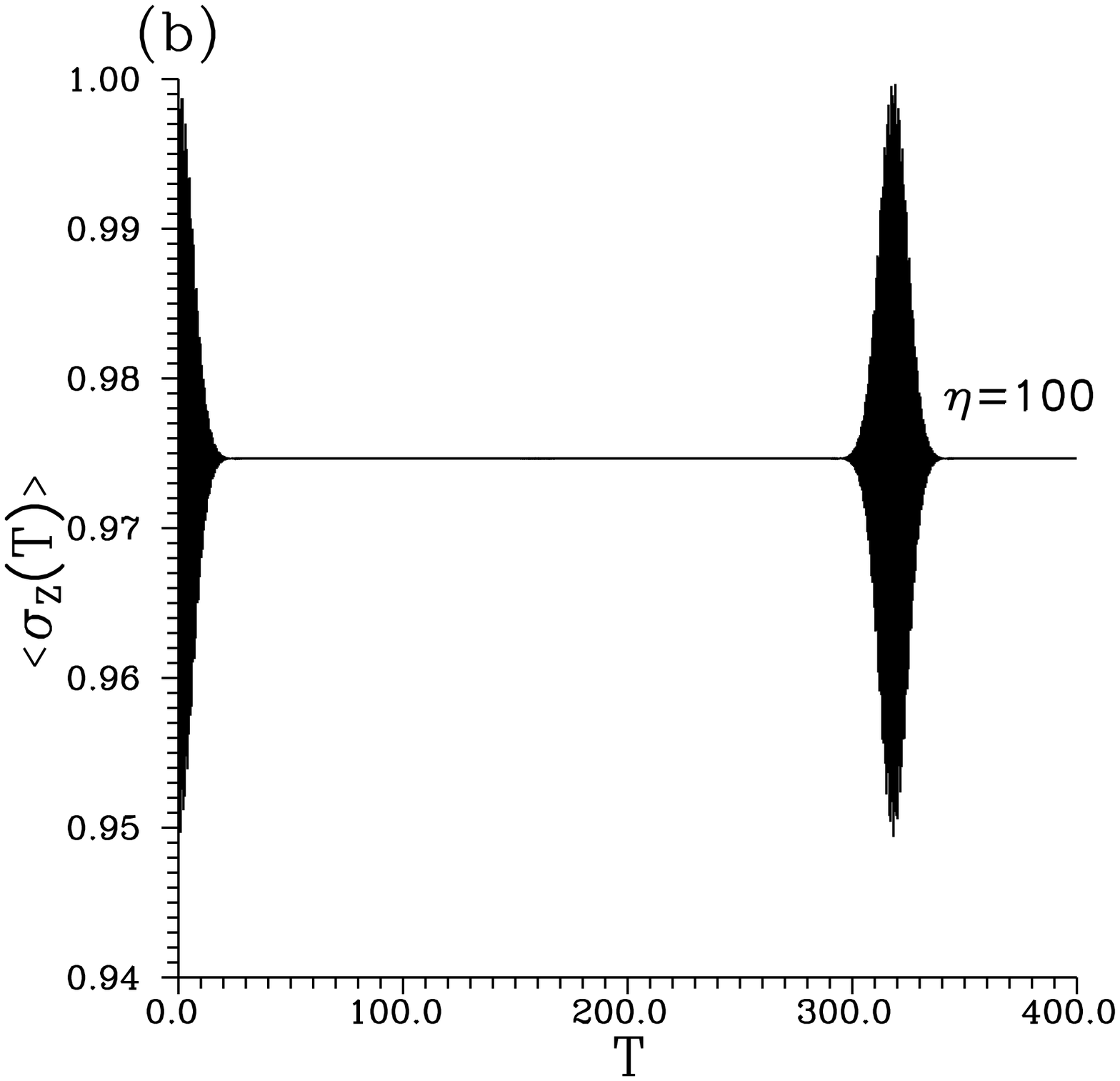}}
  \caption{
The atomic inversion $\langle\hat{\sigma}_{z}(T)\rangle$
 against the scaled time $T$ for the SJCM when the
cavity mode is  initially prepared in the  coherent state with
$(\alpha,m) =(8,1)$ and $\eta=0, 20$ (a) and $100$ (b).}
\end{figure}
%%%%%%%%%%%%%%%%%%%%%%%%%%%%%%%%%%%%%%%%%%%%%%%%%%%%%%%%%%%%
Information about (\ref{da7}) is shown in Figs. 1 and 2 for the SJCM and
IJCM, respectively, for $m=1$ and given values of the interaction parameters.
From Figs. 1 one can observe
that as the value of the detuning parameter $\eta$
increases the collapse period and the width of the revival pattern
increase. Moreover,  when the detuning is very strong
 the atomic trapping
occurs, which is close to its initial atomic state. This does not mean that
the system stays in its initial state where the IJCM
provides nonclassical squeezing for strong detuning \cite{hil}.
Also such type of the atomic trapping is quite different from that occurring
via controlling the values of the atomic relative phases \cite{zaheer}
in which $\langle\hat{\sigma}_{z}(T)\rangle\simeq 0$.
From Figs. 2, i.e. for IJCM, the RCP is remarkable, where for
$\eta=0$ the revival patterns are compact and
occur periodically with period $\pi$. This behavior is dominant when the values
of $\eta$ and $\bar{n}=\langle\hat{n}(0)\rangle$ are comparable (compare
the curves in Fig. 2(a)). For very strong value of $\eta$ the RCP
are established having features different from those for
$\eta=0$ in the sense that the revival
patterns (collapse time) become broader (greater)
than those of $\eta=0$.
Comparison between Figs. 1(b) and 2(b) shows that the atomic inversion
of the IJCM goes to trapped atomic state in a rate slower than that of the
SJCM. These facts can be partially realized from (\ref{da7}).
In this equation when  $\eta$ is very strong and
$\eta >>\bar{n}$ one can asymptotically consider
$\frac{\frac{\eta^2}{4}}{\gamma_{n,m}^{2}}\simeq 1$
(i.e., the first summation tends to unity)  and
$\frac{\frac{(n+m)!}{n!}f^2(n+m)}{\gamma_{n,m}^{2}}\leq 1$
(i.e., the absolute value of the second summation is less than
$1$ or it is close to zero). Leading that the overall behavior is close to unity.
Furthermore, when $f(.)=\sqrt{\hat{n}}$ these coefficients tend to
the specified values in rates slower than those of the case  $f(.)= 1$.
Thus the behavior associated with
the IJCM is more stable than that of the SJCM.
%%%%%%%%%%%%%%%%%%%%%%%%%%%%%%%%%%%%%%%%%%%%%%%%%%%%%%%%%%%%%%%
\begin{figure}
  {\includegraphics[width=8cm]{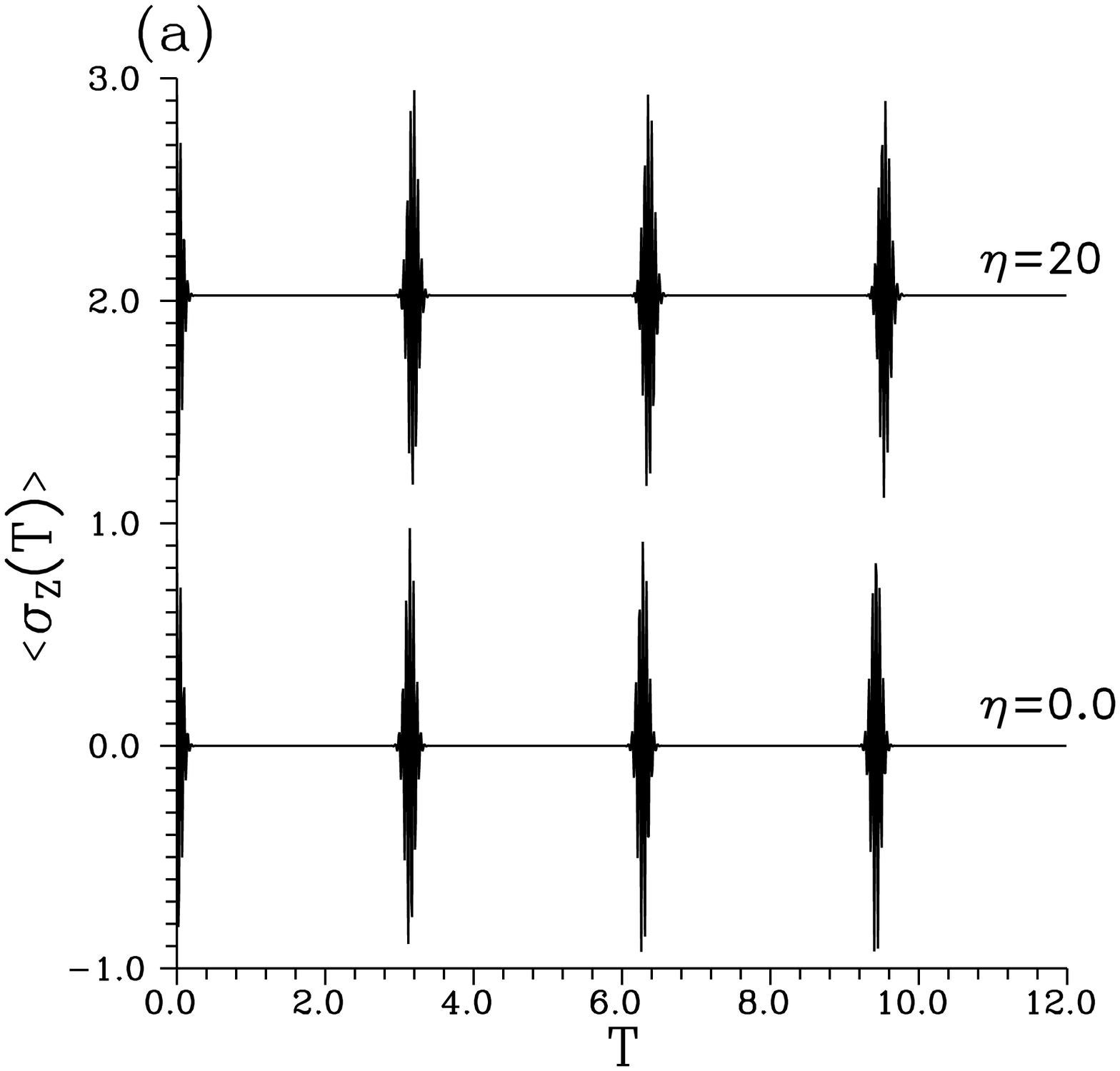}}
  {\includegraphics[width=8cm]{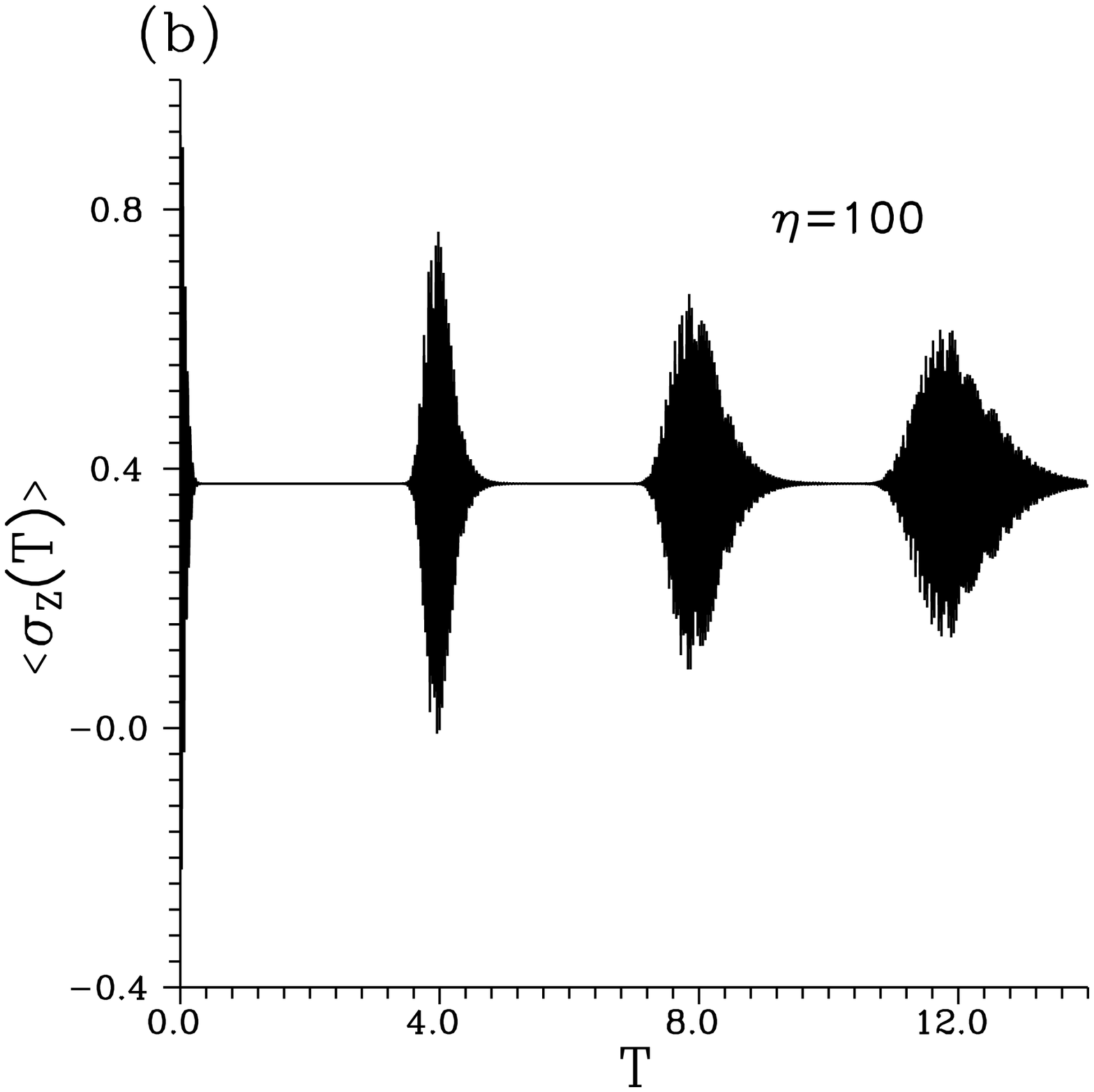}}
 \caption{ The atomic inversion
$\langle\hat{\sigma}_{z}(T)\rangle$
 against the scaled time $T$ for the IJCM when the
cavity mode is  initially prepared in the  coherent state with
$(\alpha,m) =(8,1)$ and $\eta=0, 20$ (a) and $100$ (b).}
\end{figure}
%%%%%%%%%%%%%%%%%%%%%%%%%%%%%%%%%%%%%%%%%%%%%%%%%%%%%%%%%%%%
The sensitivity of the revival time to $\eta$
can be analytically deduced. For instance, for the standard IJCM
with the radiation field prepared initially
in coherent light the revivals occur
in the $\langle\hat{\sigma}_{z}(T)\rangle$ when
neighbor terms in the sums are in phase. This can be expressed as:
\begin{eqnarray}
\begin{array}{lr}
2 T_{r}\left[ \sqrt{(\frac{\eta}{2})^2+(\langle \hat{n}(0)\rangle+1)^2}
-\sqrt{(\frac{\eta}{2})^2+ \langle \hat{n}(0)\rangle^2}\right]\simeq 2\pi,\\
\\
 T_{r}\left\{
\sqrt{(\frac{\eta}{2})^2+\langle \hat{n}(0)\rangle^2}\left[1
+ \frac{2\langle \hat{n}(0)\rangle+1}{2(\frac{\eta}{2})^2+2\langle
\hat{n}(0)\rangle^2}\right]
-\sqrt{(\frac{\eta}{2})^2+ \langle \hat{n}(0)\rangle^2}\right\}\simeq \pi,
 \label{20tq}
\end{array}
\end{eqnarray}
where $T_r$ denotes the revival time associated with the IJCM.
When $\langle \hat{n}(0)\rangle$ is large we can consider
$2\langle \hat{n}(0)\rangle+1\simeq 2\langle\hat{n}(0)\rangle$.
Consequently expression (\ref{20tq}) reduces to

\begin{equation}
 T_{r}\simeq \frac{\pi\sqrt{(\frac{\eta}{2})^2+
 \langle \hat{n}(0)\rangle^2}}{\langle \hat{n}(0)\rangle}.
 \label{20tqq}
\end{equation}
It is evident that when $\eta=0$ expression (\ref{20tqq}) reduces
 to that of the resonance case, i.e. $T_r=\pi$.
Similar procedures for SJCM with $m=1$ provide the revival time $T_r'$ as
\begin{equation}
 T_{r}'\simeq 2\pi\sqrt{(\frac{\eta}{2})^2+\langle \hat{n}(0)\rangle}.
 \label{20qq}
\end{equation}
The validity of (\ref{20qq}) and (\ref{20tqq}) can be checked for
the first revivals  of
$\eta =20, 100$ in Figs. 1 and 2.
One can also realize that when
$\eta\rightarrow \infty$ the revival times become infinity.
These results agree with those given for the nonlinear JCM in \cite{gora}.

Now--for the sake of convenience--we define
 two quadrature operators for $N$th-order squeezing  as
$\hat{X}_N=\frac{1}{2}(\hat{a}^{N}+\hat{a}^{\dagger N}),\quad
\hat{Y}_N=\frac{1}{2i}(\hat{a}^{N}-\hat{a}^{\dagger N})$,
where $N$ is a positive integer.
These quadratures satisfy the commutation rule
$[ \hat{X},\hat{Y}] =\frac{i\hat{C}_N}{2}$,
where $\hat{C}_N=
\hat{a}^{N}\hat{a}^{\dagger N}- \hat{a}^{\dagger
N}\hat{a}^{N}$.
Thus the uncertainty relation is
$\langle (\triangle\hat{X}(T))^{2}\rangle
  \langle (\triangle\hat{Y}(T))^{2}\rangle \geq \frac{|\langle
  \hat{C}_N\rangle|^2  }{16}$.
The squeezing factors associated with  $\hat{X}_N$ and  $\hat{Y}_N$
can be respectively expressed as \cite{hig}:

\begin{eqnarray}
\begin{array}{lr}
S_N(T)
= \langle\hat{a}^{\dagger N}(T)\hat{a}^{N}(T)\rangle+{\rm Re}
\langle\hat{a}^{2N}(T)\rangle-2({\rm
Re}\langle\hat{a}^{N}(T)\rangle)^{2},
\\
\\
Q_N(T)=
 \langle\hat{a}^{\dagger N}(T)\hat{a}^{N}(T)\rangle
-{\rm Re}\langle\hat{a}^{2N}(T)\rangle-2({\rm
Im}\langle\hat{a}^{N}(T)\rangle)^{2}.
\label{s14}
\end{array}
\end{eqnarray}

Here we shed briefly the light on the natural approach for the IJCM
since it is similar to that of the SJCM \cite{faisal2}.
In this approach the squeezing factors
provide direct information on the atomic inversion. This is based on the
fact that
 $\hat{F}_1$ is a constant of
motion, i.e.  the quantities $\langle\hat{\sigma}_{z}(T)\rangle$ and
$\langle\hat{a}^{\dagger }(T)\hat{a}(T)\rangle$
 carry information on each other.
This approach is established for particular type of initial states
of the radiation field, which verify, e.g. for $N=1$ normal squeezing,
the conditions

\begin{equation}
\langle\hat{a}(T)\rangle=0,\qquad \langle\hat{a}^{2}(T)\rangle=0
\label{s15}
\end{equation}
simultaneously.
This can occur when the mode
is initially prepared in three-photon states \cite{gener},
 four-photon states \cite{lyn},.. etc.
In this case, similar to the SJCM \cite{fais1}, one can easily prove that
\begin{equation}
\langle\hat{\sigma}_{z}(T)\rangle= 2\langle\hat{a}^{\dagger
}(0)\hat{a}(0)\rangle +1 -2S_{1}(T). \label{ss15}
\end{equation}
Expression (\ref{ss15}) shows that the atomic inversion can be readout
from the quadrature squeezing.
We have to stress that the expression (\ref{ss15}) is the same for
resonance and off-resonance cases.
Moreover, expression of the higher-order squeezing
for IJCM can be also derived similar to that in \cite{faisal2}.

In the following sections we use the relations given in the present section to
investigate numerically how within the multiphoton IJCM
 the quadratures squeezing exhibit the RCP
similar to that involved in the evolution of the atomic inversion of
the standard IJCM with initial coherent light.
We discuss to which value of the initial mean-photon number
the quadrature squeezing gives the most appropriate information on the atomic inversion.
Also we investigate the influence of the detuning parameter
on the under consideration phenomenon for SJCM and IJCM.

%%%%%%%%%%%%%%%%%%%%%%%%%%%%%%%%%%%%%%%%%%%%%%%%%%%%%%%%%%%%%%%%%%%%%%%
\section{Resonance case}
%%%%%%%%%%%%%%%%%%%%%%%%%%%%%%%%%%%%%%%%%%%%%%%%%%%%%%%%%%%%%%%%%%%%%%%%
In this section we investigate the possibility to obtain RCP from the
second-order quadrature squeezing of the  $m$th-photon ($m>1$)
IJCM for  $\eta=0$ similar to that of the standard
IJCM (, i.e. $\langle \hat{\sigma}_{z}(T)\rangle_{m=1}$)
 when the field is initially in the coherent state.
According to the line given in \cite{faisal2} the
RCP can occur only in $Q_{1}(T)$ and the quantity
${\rm Re} \langle \hat{a}^{2}(T)\rangle$ is responsible for
such behavior.
Thus  we give a closer look at the expression of the $\langle
\hat{a}^{2}(T)\rangle$, which from (\ref{8}) can be evaluated as:

\begin{eqnarray}
\begin{array}{lr}
\langle \hat{a}^{2}(T)\rangle=
\alpha^2\sum\limits_{n=0}^{\infty}
P(n)
\Bigl\{
\cos(T\gamma_{n+2,m})\cos(T\gamma_{n,m})\\
\\
+ \sqrt{\frac{(n+m+1)(n+m+2)}{(n+1)(n+2)}}
\sin(T\gamma_{n+2,m})\sin(T\gamma_{n,m})\Bigr\}.
\label{s14s}
\end{array}
\end{eqnarray}
In the  strong-intensity regime, i.e. $\bar{n}=|\alpha|^2 >>1$, and
 finite values of the transition parameter $m$,
 we can apply the harmonic approximation technique
\cite{eber}:
for Poissonian
  photon-number distribution  such as that of the coherent light
 the terms  contributing
 effectively to the summation in (\ref{s14s}) are those for which
 $ n\simeq\bar{n}$. Consequently
 the  square root in the second line of (\ref{s14s})
 tends to unity and (\ref{s14s}) reduces to

\begin{equation}
\langle \hat{a}^{2}(T)\rangle\simeq
\bar{n}\sum\limits_{n=0}^{\infty}P(n)
\cos[T(\gamma_{n+2,m}-\gamma_{n,m})].
\label{1s4s}
\end{equation}
Comparison between (\ref{da7}) (for $\eta =0$ and $m=1$)
  and (\ref{1s4s}) shows that the two
expressions can exhibit  similar dynamical behavior only
when the arguments
of cosines in the two expressions are comparable. Therefore, we seek
 the proportionality factor $\mu_1$, say, which can play this role.
 This factor can be evaluated from the
following expression:
\begin{equation}
\mu_1=\frac{
\gamma_{n+2,m}-\gamma_{n,m}}{2(n+1)}.
\label{pro1}
\end{equation}
Expression (\ref{pro1}) can be re-expressed  as
\begin{eqnarray}
\begin{array}{lr}
\mu_1=\frac{1}{2(n+1)}\sqrt{\frac{(n+m)!}{(n+2)!}}
\Bigl\{\sqrt{(n+m+2)^2(n+m+1)}-\sqrt{(n+m)(n+1)(n+2)}
\Bigr\}\\
\\
=\frac{n^{\frac{m-7}{2}}}{2(1+\frac{1}{n})\sqrt{(1+\frac{1}{n})(1+\frac{2}{n})}}
\left[\prod\limits_{j=0}^{m}(1+\frac{m-j}{n})\right]^{\frac{1}{2}}
\\
\\
\times
\frac{\left[2n^2(m+1)+(3m^2+7m+6)n+(m+2)^2(1+m)-2m\right]}
{[(1+\frac{m+2}{n})
\sqrt{1+\frac{m+1}{n}}
+\sqrt{(1+\frac{m}{n})(1+\frac{1}{n})(1+\frac{2}{n})}]}.
\label{pro2}
\end{array}
\end{eqnarray}
%%%%%%%%%%%%%%%%%%%%%%%%%%%%%%%%%%%%%%%%%%%%%%%%%%%%%%%%%%%%%%%
\begin{figure}
{\includegraphics[width=8cm]{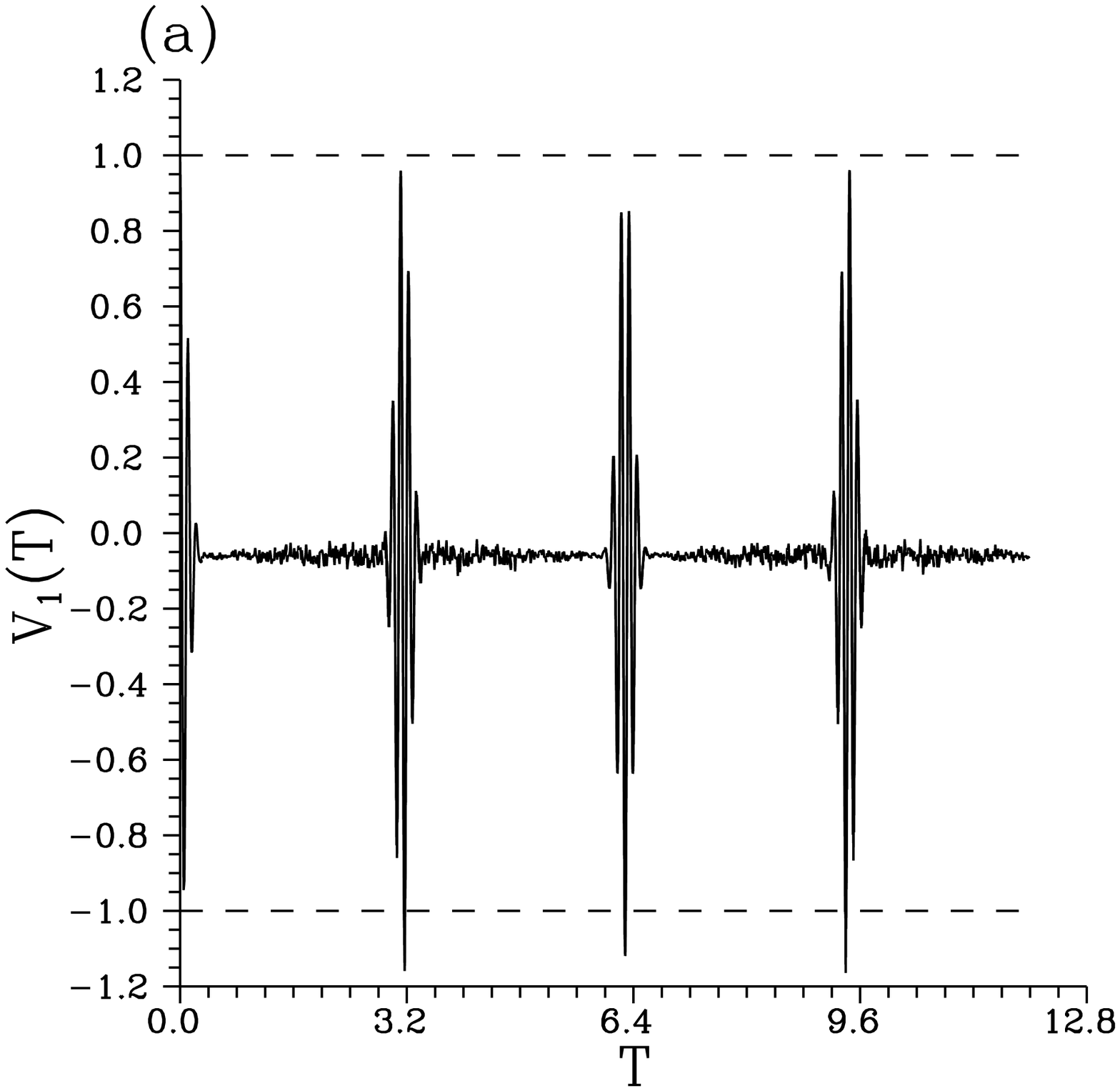}}
{\includegraphics[width=8cm]{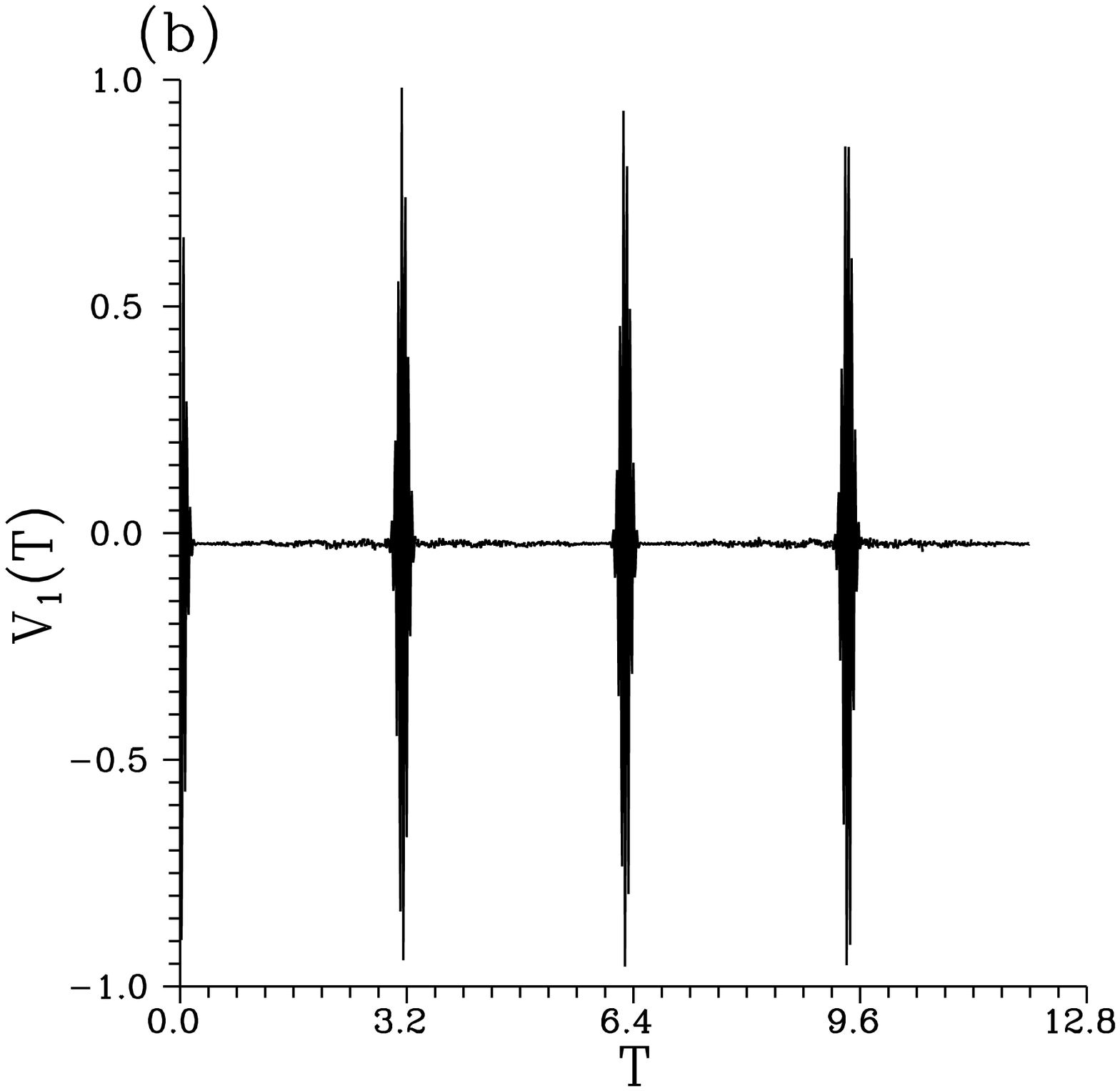}}
\caption{ The rescaled
squeezing factor $V_1(T)$ for the IJCM when $m=3$, $\alpha=5$ (a)
and $8$ (b).}
\end{figure}
%%%%%%%%%%%%%%%%%%%%%%%%%%%%%%%%%%%%%%%%%%%%%%%%%%%%%%%%%%%%
In the strong-intensity regime, where the harmonic approximation can be
applied (, i.e. $\epsilon/n\rightarrow 0$, where $\epsilon$ is an arbitrary
finite number) the equation (\ref{pro2}) reduces to
\begin{equation}
\mu_1\simeq \frac{1}{4}\Bigl\{ 2(m+1)\bar{n}^{\frac{m-3}{2}}
+(3m^2+7m+6)\bar{n}^{\frac{m-5}{2}}
+[(m+2)^2(1+m)-2m]\bar{n}^{\frac{m-7}{2}}\Bigr\}. \label{pro3}
\end{equation}
From (\ref{pro3}) the RCP can occur
in the evolution of the $Q_1(T)$ only when $m=3$, which is similar to that of
SJCM \cite{faisal2}. In this case the proportionality factor is
$\mu_1=2$. Furthermore, one can use the same procedures (for complete
details the reader can consult \cite{fais2,{fais1}}) for evaluating the proportionality
factor of the $N$th-order squeezing, which is $\mu_1=2N$.
For the sake of generalization we write down
 the $N$th-order rescaled squeezing factor as

\begin{equation}
V_{N}(T)=\frac{\langle\hat{n}(0)\rangle^N
-Q(\frac{bT}{N})}
{\langle\hat{n}(0)\rangle^N}, \label{29q}
\end{equation}
where the parameter $b$ takes on $2/3$ for $f(.)=1$ \cite{faisal2,{fais1}}
 and $1/2$ for
$f(.)=\sqrt{\hat{n}}$, which is the case under consideration.
 For normal squeezing and $f(.)=\sqrt{\hat{n}}$ we have plotted $V_1(T)$ for
$\alpha=5$ (a) and $\alpha=8$ (b) against the scaled time $T$ in Figs.
3(a) and (b), respectively.
Comparison between the curve associated with $\eta=0$ in Fig. 2(a) with
those presented in Figs. 3 demonstrate our conclusion: the rescaled
squeezing factor (\ref{29q}) can give information on the atomic inversion of the
standard IJCM. On the other hand, we can see that both curves in Figs. 3
provide information on the atomic inversion, however,
 Fig. 3(b) exhibits better information
than that of the Fig. 3(a) (compare the shape of the collapse regions
and the amplitude of the revival patterns).  This leads to the following question: which values of the
initial mean-photon number can give perfect information on the atomic inversion?
The answer for this question can be easily realized, e.g. for normal squeezing,
 from (\ref{pro3}). More precisely,
the technique works well when the value of the second term in the rectangular
brackets is close to zero. This can be expressed as:
%%%%%%%%%%%%%%%%%%%%%%%%%%%%%%%%%%%%%%%%%%%%%%%%%%%%%%%%%%%%%%%
\begin{figure}
  {\includegraphics[width=8cm]{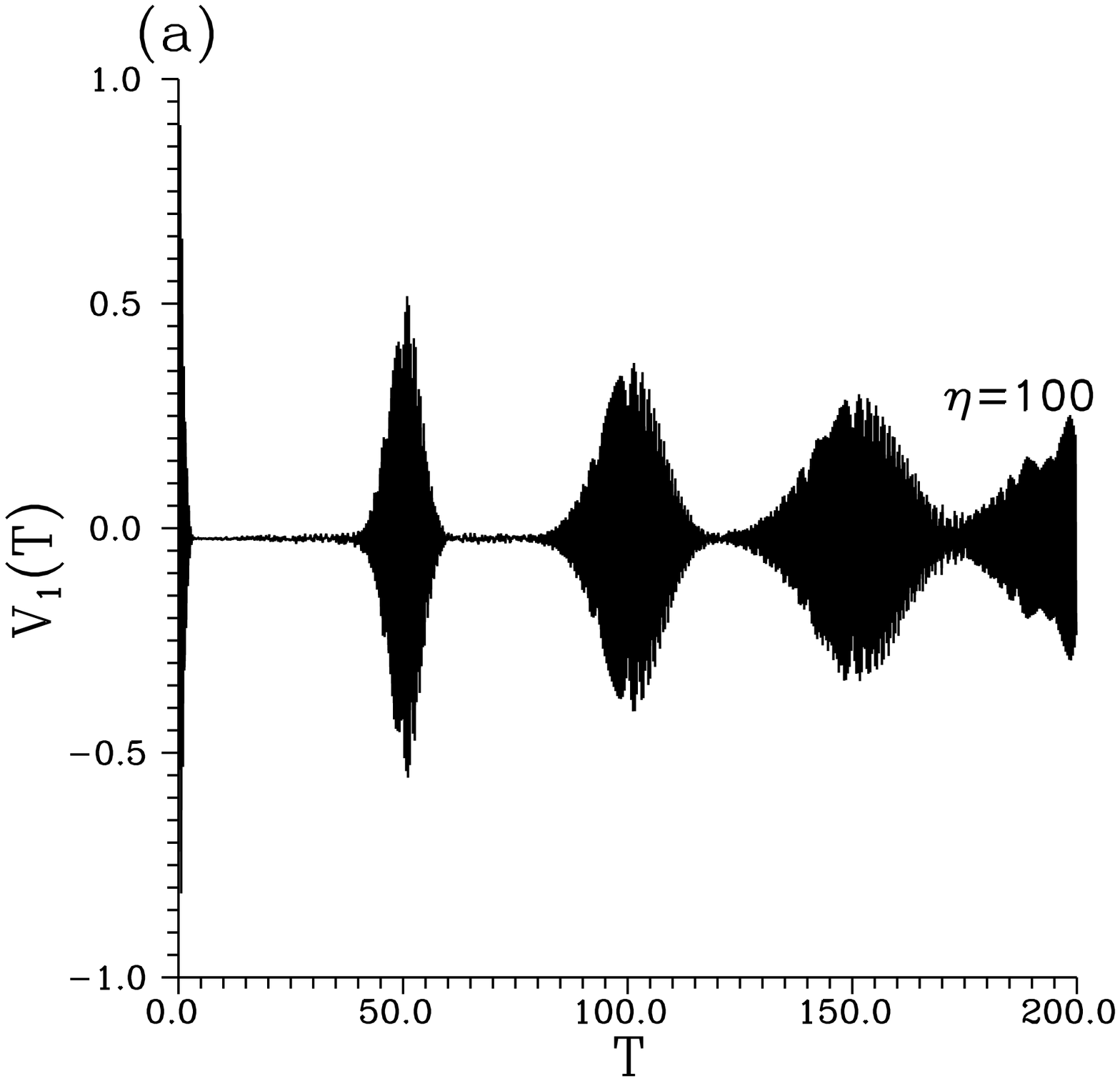}}
  {\includegraphics[width=8cm]{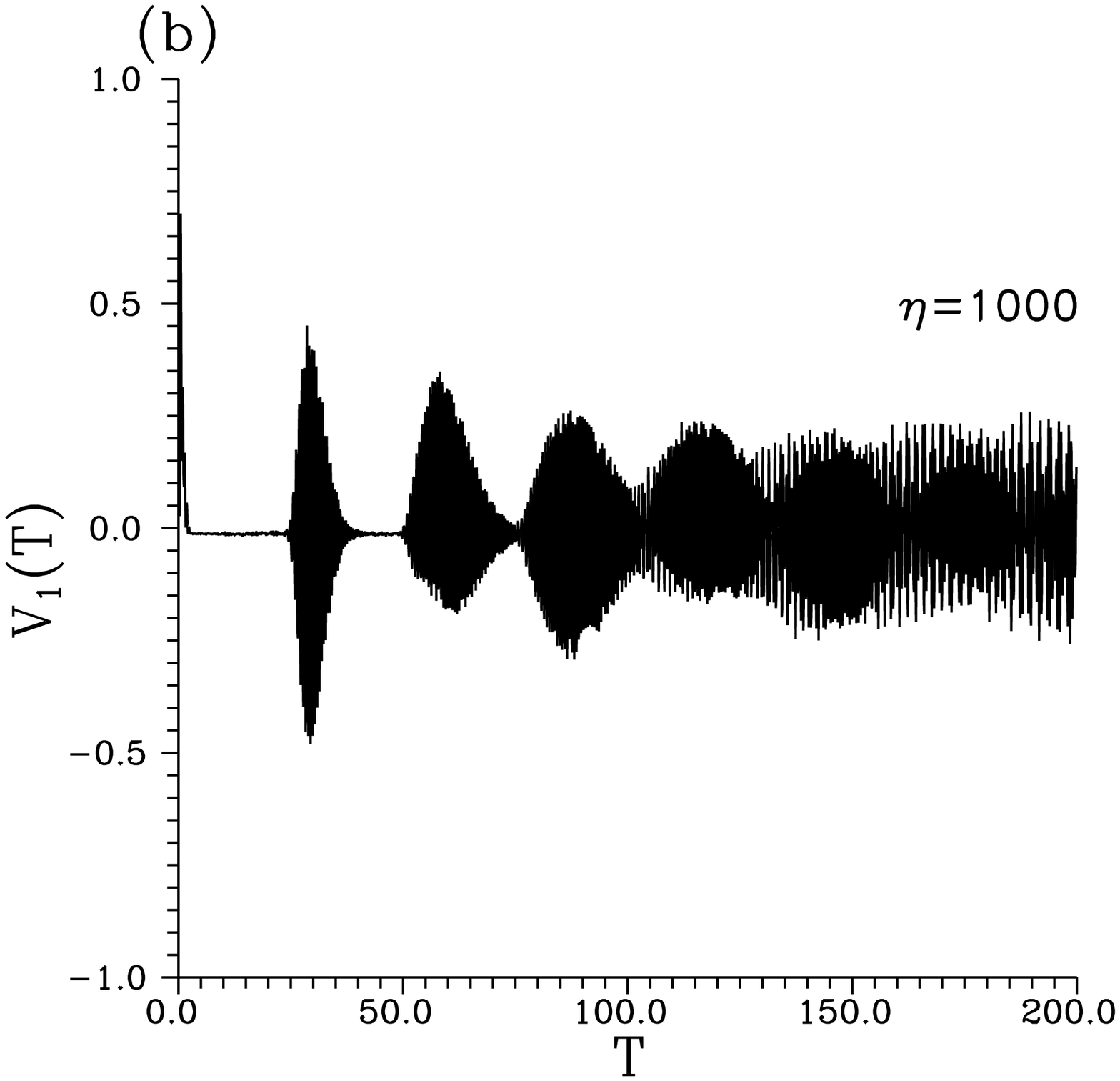}}
\caption{ The rescaled squeezing factor $V_1(T)$ for the SJCM when
$(\alpha,m)=(8,3)$, $\eta =100$ (a) and $1000$ (b).}
\end{figure}
%%%%%%%%%%%%%%%%%%%%%%%%%%%%%%%%%%%%%%%%%%%%%%%%%%%%%%%%%%%%

\begin{equation}
(3m^2+7m+6)\bar{n}^{\frac{m-5}{2}}<1. \label{staf}
\end{equation}
For $m=3$ the inequality (\ref{staf}) reduces to $\bar{n}>54$.
This agrees with the information shown in Figs. 3.
As the value of the squeezing-order $N$ increases the value of the
initial mean-photon number required for giving better information on
$\langle \hat{\sigma}_{z}(T)\rangle_{m=1}$ has to be increased, e.g.
for the amplitude-squared squeezing we have to use
$\langle\hat{n}(0)\rangle>226$ (we have checked this fact).
On the other hand, for SJCM, i.e. $f(.)=1$, the $\langle\hat{n}(0)\rangle>18$ and $144$
for the normal and amplitude-squared squeezing, respectively.
In conclusion, for strong initial mean-photon number  the formula
(\ref{29q}) works regardless of the condition (\ref{staf}) where
the overall behaviors of the rescaled squeezing factor and
atomic inversion are similar. Hoever, to obtain typical information  from
these two quantities such type of  condition has to be fulfilled.

%%%%%%%%%%%%%%%%%%%%%%%%%%%%%%%%%%%%%%%%%%%%%%%%%%%%%%%%%%%%%%%%%%%%%%%%
%%%%%%%%%%%%%%%%%%%%%%%%%%%%%%%%%%%%%%%%%%%%%%%%%%%%%%%%%%%%%%%%%%%%%%%%
\section{Off-resonance case}
%%%%%%%%%%%%%%%%%%%%%%%%%%%%%%%%%%%%%%%%%%%%%%%%%%%%%%%%%%%%%%%%%%%%%%%%%
%%%%%%%%%%%%%%%%%%%%%%%%%%%%%%%%%%%%%%%%%%%%%%%%%%%%%%%%%%%%%%%%%%%%%%%%%
In this section we discuss the off-resonance case $\eta\neq 0$ for the multiphoton
 SJCM and IJCM in the framework of the numerical approach.
It is worth reminding that the $\langle \hat{\sigma}_{z}(T)\rangle$
is sensitive to the value of the detuning parameter.
We restrict the discussion to the normal squeezing, where the higher-order squeezing
can be similarly understood.
There are some facts we would like to address here.
For $\eta\neq 0$ we have
 ${\rm Re} \langle \hat{a}(T)\rangle\neq 0$ and
 ${\rm Im} \langle \hat{a}(T)\rangle \neq 0$, which likely indicates that
 the technique does not work correctly.
Nevertheless, we have noted that
the values of ${\rm Im} \langle \hat{a}(T)\rangle$ are considerably small
compared to the other terms in the squeezing factor. This
means that if the squeezing factors exhibit
information about $\langle \hat{\sigma}_{z}(T)\rangle$
this probably occur in $Q_1(T)$. In this case the
quantity ${\rm Re} \langle \hat{a}^2(T)\rangle$ is responsible for
this. This behavior is typical to that
of the resonance case (cf. section 3).
As there is a difficulty to derive the rescaled squeezing
factor for $\eta \neq 0$
 we investigate the influence of $\eta$ on the formula (\ref{29q}).
We have noted that this formula for
$\eta\neq 0$ provides RCP typical to that of
$\eta = 0$, however,
for very strong values of $\eta$ the RCP is different
from that of the
$\langle \hat{\sigma}_{z}(T)\rangle_{m=1}$.
These facts are shown
in Figs. 4 and 5 for given values of the parameters.
For SJCM comparison between Figs. 1(a), 1(b) and Fig. 4(a) demonstrate the
main conclusion. From Fig. 4(b) where $\eta$ is very strong the revival
patterns are increased, blurred and overlapped for large time interaction.
This is in contrast with the corresponding case of the
$\langle \hat{\sigma}_{z}(T)\rangle_{m=1}$.
Similar remarks may be quoted for IJCM (see Figs. 5).
Additionally, the RCP involved in (\ref{29q})
for IJCM is much stable than that for SJCM, i.e. it needs values of
$\eta$ greater than those for SJCM to be deviated
from that of the  atomic inversion.

%%%%%%%%%%%%%%%%%%%%%%%%%%%%%%%%%%%%%%%%%%%%%%%%%%%%%%%%%%%%%%%
\begin{figure}
 {\includegraphics[width=8cm]{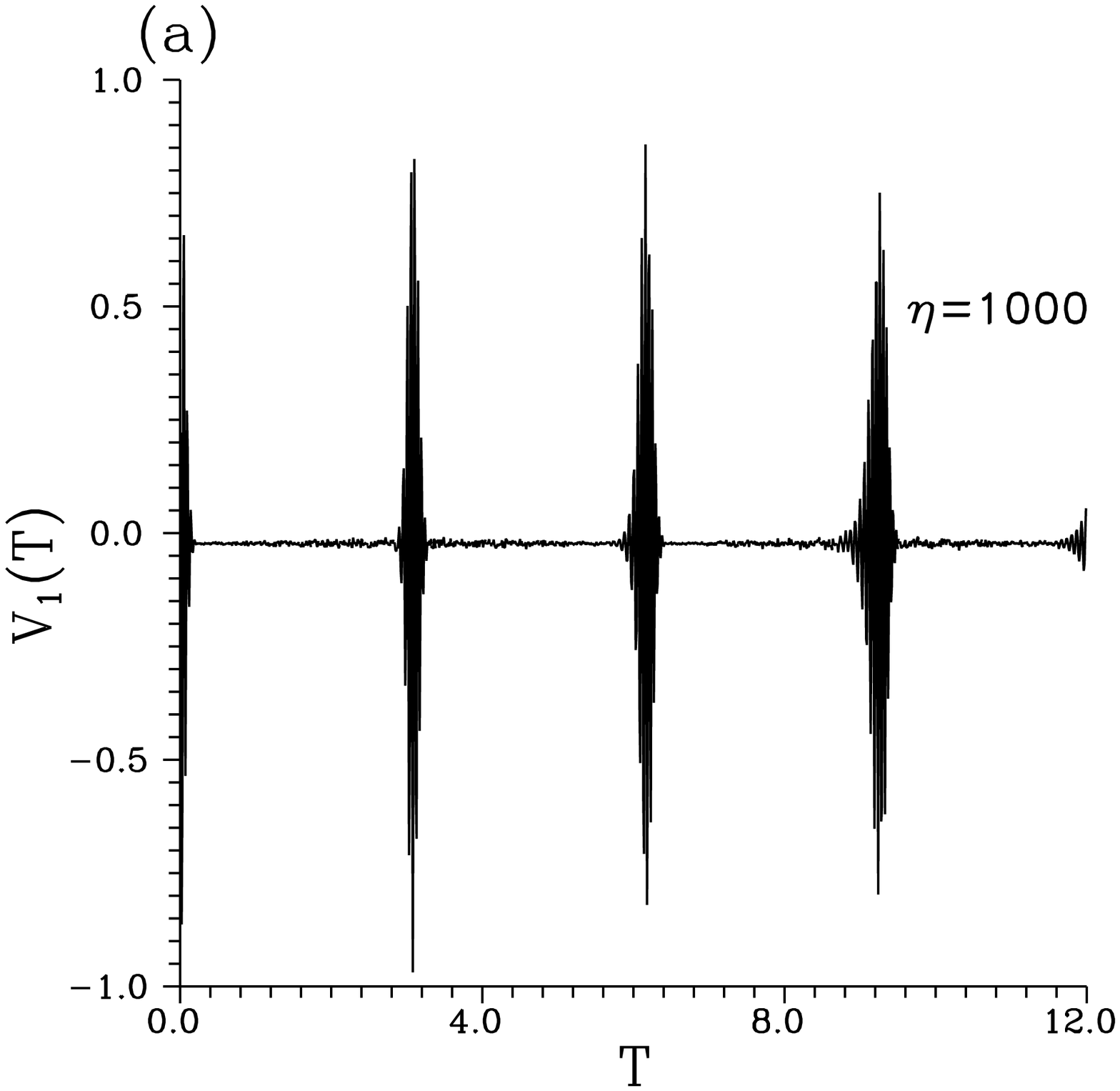}}
 {\includegraphics[width=8cm]{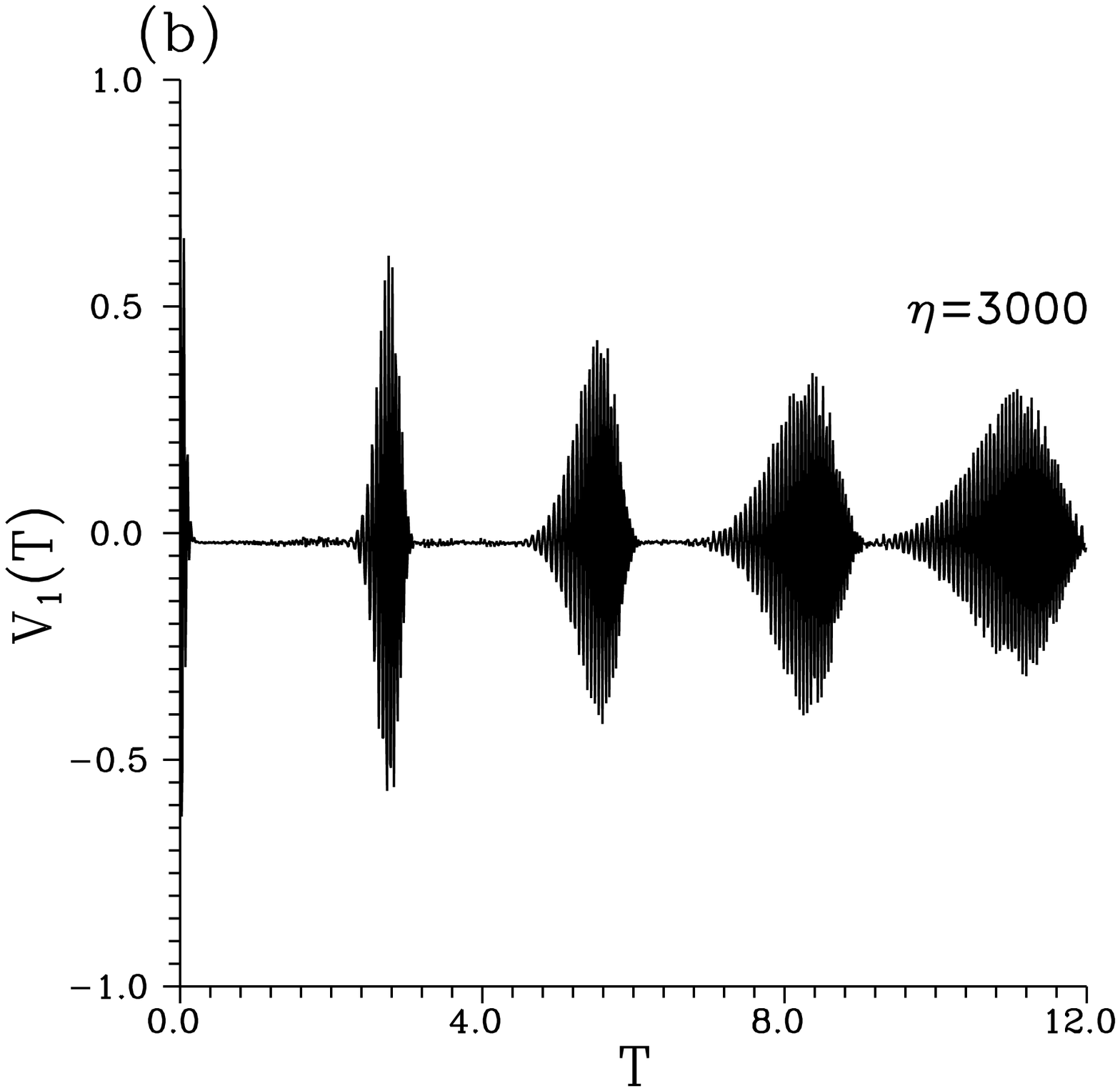}}
\caption{The rescaled squeezing factor $V_1(T)$ for the IJCM when
$(\alpha,m)=(8,3)$, $\eta =1000$ (a) and $3000$ (b).}
\end{figure}
%%%%%%%%%%%%%%%%%%%%%%%%%%%%%%%%%%%%%%%%%%%%%%%%%%%%%%%%%%%%
Now we partially explain the results shown in Figs. 4 and 5 as follows.
The quantity ${\rm Re} \langle \hat{a}^2(T)\rangle$ plays the essential
role in the behavior of the rescaled squeezing factor, which can be evaluated
for the state (\ref{8}) as:
\begin{eqnarray}
\begin{array}{lr}
{\rm Re}\langle \hat{a}^{2}(T)\rangle=
\alpha^2\sum\limits_{n=0}^{\infty}P(n)
\Bigl\{\cos(T\gamma_{n+2,m})\cos(T\gamma_{n,m})\\
\\
+\left[\frac{\frac{\eta^2}{4}+\frac{(n+m+2)!}{(n+2)!}f(n+m)f(n+m+2)}
{\gamma_{n+2,m}\gamma_{n,m}} \right]
\sin(T\gamma_{n+2,m})\sin(T\gamma_{n,m})\Bigr\}.
\label{ss}
\end{array}
\end{eqnarray}
In the strong-intensity regime and finite value of $m$ the rectangular
brackets in the second line of (\ref{ss}) tends to unity and we arrive
at
\begin{equation}
{\rm Re}\langle \hat{a}^{2}(T)\rangle=
\alpha^2\sum\limits_{n=0}^{\infty}
P(n)
\cos(T\Omega_n),
\label{ss1}
\end{equation}
where the generalized Rabi frequency $\Omega_n$ takes the form
\begin{equation}
\Omega_n=
\sqrt{\frac{\eta^2}{4}+\frac{(n+m+2)!}{(n+2)!}f^{2}(n+m+2)}
-\sqrt{\frac{\eta^2}{4}+\frac{(n+m)!}{n!}f^{2}(n+m)}.
\label{ss2}
\end{equation}
The behavior of the $V_1(T)$ is a direct consequence form the "evolution"
of the generalized
 Rabi frequency $\Omega_n$ in the range of $n$ for which $P(n)$ has an
effective contribution. Therefore, for IJCM we plot $\Omega_n$ and $P(n)$ against
$n$ in Fig. 6 for values of parameters as those given for Figs. 5.
For the sake of comparison between these two quantities we rescaled $\Omega_n$
by $1/6000$.
%%%%%%%%%%%%%%%%%%%%%%%%%%%%%%%%%%%%%%%%%%%%%%%%%%%%%%%%%%%%%%%
\begin{figure}
  {\includegraphics[width=8cm]{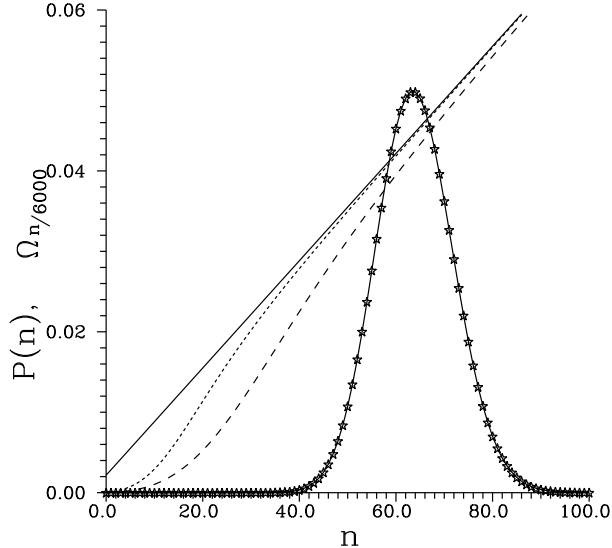}}
\caption{ The generalized Rabi frequency $\Omega_n$ and the
photon-number distribution $P(n)$ (star-centered curve) against
$n$ for IJCM when $(\alpha,m)=(8,3)$. For $\Omega_n$ solid,
short-dashed and long-dashed curves are given respectively for
$\eta =0, 1000$ and $3000$.}
\end{figure}
%%%%%%%%%%%%%%%%%%%%%%%%%%%%%%%%%%%%%%%%%%%%%%%%%%%%%%%%%%%%
From this figure one can observe that for $\eta=0$ the generalized Rabi frequency
$\Omega_n$ is linear, i.e. straightline curve, however,
when $\eta\neq 0$ nonlinearity occurs (see short-dashed and long-dashed curves).
In all cases the curves are growing function of $n$.
Most importantly, for $\eta =1000$ the generalized Rabi frequency in the range
of $P(n)\neq 0$ is almost equal to that of $\eta =0$.
This makes the behavior
of $V_1(T)$ for these two cases are almost the same.
Comparison between the short-dashed and long-dashed curves shows that
when $\eta$ is very large the
deviation form the resonance case is established. This agrees
with information shown in Figs. 5(a) and
(b).  Similar treatment can be given for SJCM.
%%%%%%%%%%%%%%%%%%%%%%%%%%%%%%%%%%%%%%%%%%%%%%%%%%%%%%%%%%%%%%%%%%%%%%%
\section{Conclusions}
%%%%%%%%%%%%%%%%%%%%%%%%%%%%%%%%%%%%%%%%%%%%%%%%%%%%%%%%%%%%%%%%%%%%%%%
In this paper we have discussed the possibility of including
the squeezing factors of the  multiphoton
IJCM information on the atomic inversion of the standard IJCM.
We have discussed the restriction on the values of
the initial mean-photon number, which can
give the most appropriate information about the atomic inversion.
For the natural approach we have briefly shown that for
particular type of initial states the squeezing factor for the off-resonance
IJCM is similar
to that of the resonance SJCM \cite{faisal2}. In this approach the
squeezing factors provide complete information on the atomic inversion.
For the resonance IJCM we have shown that
the rescaled squeezing factor of the three-photon transition IJCM
can give information on the atomic inversion of the standard IJCM.
On the other hand, for the off-resonance IJCM and SJCM
 we have shown that the rescaled squeezing factor gives always
$\langle \hat{\sigma}_{z}(T)\rangle_{m=1}$ of the $\Delta=0$
for the finite values of the detuning parameter, however,
for the large values of $\Delta$ it exhibits RCP different from that of
the atomic inversion. We have explained such behavior
by comparing the evolution of the generalized Rabi frequency for
 different values of $\Delta$.

%%%%%%%%%%%%%%%%%%%%%%%%%%%%%%%%%%%%%%%%%%%%%%%%%%%%%%%%%%%%%%%%%%%%%%%%%%%%%%%%
%%%%%%%%%%%%%%%%%%%%%%%%%%%%%%%%%%%%%%%%%%%%%%%%%%%%%%%%%%%%%%%%%%%%%%%%%%%%%%%%
\section*{References}
%%%%%%%%%%%%%%%%%%%%%%%%%%%%%%%%%%%%%%%%%%%%%%%%%%%%%%%%%%%%%%%%%%%%%%%%%%%%%%%%
%%%%%%%%%%%%%%%%%%%%%%%%%%%%%%%%%%%%%%%%%%%%%%%%%%%%%%%%%%%%%%%%%%%%%%%%%%%%%%%%

\end{document}